# Graphene Resonant Pressure Sensor with Ultrahigh Responsivity

*Swapnil More\* and Akshay Naik*

Centre for Nano Science and Engineering, Indian Institute of Science, Bangalore 560012

**Abstract:** Graphene has good mechanical properties including large Young's modulus, making it ideal for many resonant sensing applications. Nonetheless, the development of graphene based sensors has been limited due to difficulties in fabrication, encapsulation, and packaging. Here we report a graphene nanoresonator based resonant pressure sensor. The graphene nano resonator is fabricated on a thin silicon diaphragm that deforms due to pressure differential across it. The deformation-induced strain change results in a resonance frequency shift of the graphene nano resonator. The pressure sensing experiments demonstrate a record high responsivity of 20Hz/Pa with a resolution of 90Pa. The resolution of the sensing scheme is 0.003% of the full-scale range of the pressure sensor. This exceptional performance is attributed to two factors: maintaining a high-quality vacuum environment for the nanoresonator and introducing stimuli through a thin silicon diaphragm. The proposed pressure sensor design provides flexibility to adjust responsivity and range as needed. The fabrication method is simple and has the potential to be integrated with standard CMOS fabrication. The innovative substrate packaging allows the coupling of the resonator's strain with pressure.

Advances in micromachining technologies have enabled the miniaturization of traditional pressure sensors. Micro-scale pressure sensors not only offer economy of scale but also have better performance in terms of responsivity, sensitivity and range. Most micro-pressure sensors employ a thin silicon diaphragm that deforms when subjected to a pressure difference across it. The strain change produced by such deformation can be measured by piezoresistive [1] [2] [3] [4], capacitive [5] [6] [7], or optical techniques [8] [9] [10] [11] or by measuring the change in resonance frequency of a resonator fabricated on the diaphragm [12] [13] [14].

Resonant pressure sensing has emerged as a highly stable and accurate technique over the other sensing schemes. The typical resonator of choice for such a pressure sensor is a silicon structure, which can be easily carved out of the bulk silicon. The performance of silicon resonator-based resonant pressure sensors is limited by the mechanical properties of silicon. Typical responsivities of silicon resonator-based resonant pressure sensors are in the range of a few 100Hz/kPa [12] [13] [14]. Since the resonance frequency is proportional resonator's Young's modulus, the simplest way to improve responsivity is to use a high Young's modulus material for the resonator. 2D materials-based resonators have ultra-high responsivity because of their superior mechanical properties [15] [16] [17] [18] [19] [20]. A graphene nanoresonator is predicted to have responsivity two orders of magnitude higher than the conventional resonant pressure sensors [21]. The primary reason for this high responsivity is the high Young's modulus of graphene, which is about 1 terapascals [22].

A major hurdle in reliable graphene nanoresonator-based pressure sensors has been the fabrication and packaging scheme. The graphene nanoresonator needs to be fabricated on a silicon diaphragm, and there needs to be an overall package that facilitates a pressure difference across the silicon diaphragm. In our previous work, we had presented a practical fabrication scheme and a sensor package to realize such a resonant strain transducer [23]. In this study, we report on the performance of graphene resonant pressure sensors fabricated and packaged using the proposed scheme. We demonstrate a graphene resonant pressure sensor with a responsivity of 20Hz/Pa, which, to our knowledge, is the highest so far for a resonant pressure sensor. The devices are tested over a range of 270kPa. The data from our experiments underscore the importance of graphene NEMS as an ultra-responsive resonant pressure sensor. The small footprint of the resonator offers a promising outlook for the large-scale integration of graphene-based resonant pressure sensors.

The PCB and device chip assembly are at the heart of the sensor package (section S1 supplementary material). The graphene nanoresonators are fabricated on a $9mm \times 9mm$ Si/SiO$_2$ substrate using the method reported in our previous work [23]. Figure 1(a) and (b) show the SEM images of the graphene nanoresonators. The Raman spectrum of the graphene shows that the graphene is trilayer (Figure 1(c)). The substrate has a diaphragm of $4mm$ radius and $150 \mu m$ thickness. The backside of the chip can be seen in Figure 1(d). The location and orientation of the graphene nanoresonators on the silicon diaphragm are shown in Figure 1(e). The red dot in Figure 1(e) shows the location of the diaphragm center. Device 1 (Dev-1) is 50 $\mu m$ and device 2 (Dev-2) is 67 $\mu m$ away from the diaphragm centre. Both resonators are aligned in the radial direction with respect to the diaphragm center. This was achieved by patterning the graphene layer in the required shape and orientation using $O_2$ plasma etching. The alignment of the resonators along the radial direction ensures that the entire strain experienced by the 2D NEMS is due to the radial component of the substrate's strain [23]. We assume the tangential component of the diaphragm strain does not contribute to the device strain, thus simplifying subsequent analysis. The chip is then attached to a PCB with connectors for input and output signals (figure S1 supplementary material). The pressure difference between the front and back sides of the PCB deforms the silicon diaphragm leading to the strain change at the graphene nanoresonators fabricated on the front side of the diaphragm.

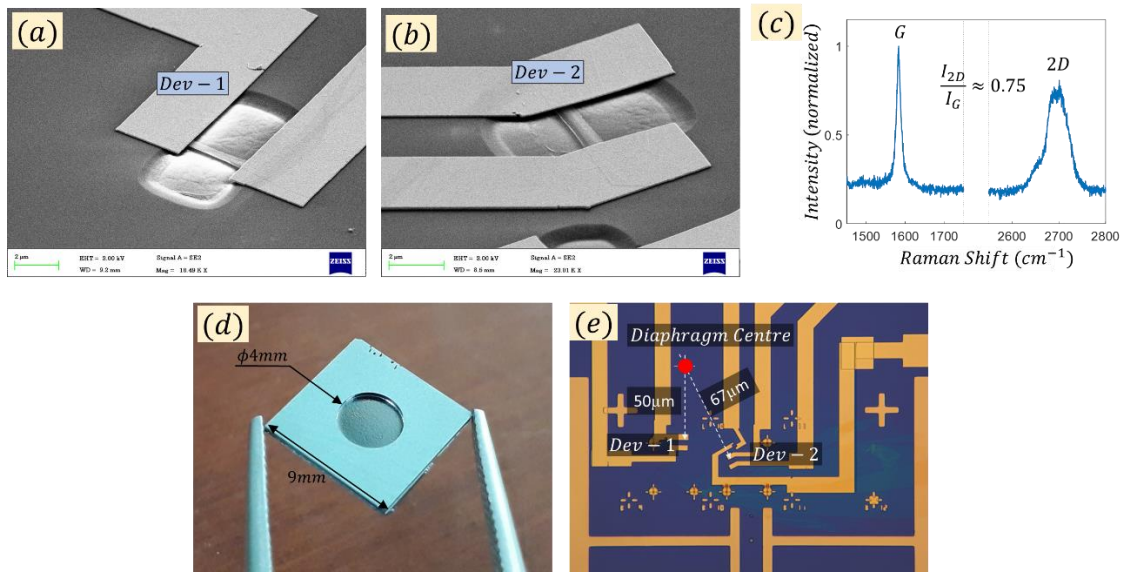

Figure 1: (a) and (b) SEM images of Dev-1 and Dev-2. (c) The Raman spectrum of the graphene used to fabricate these devices. The location, peak shape and relative intensity of the 2D peak

strongly suggest the graphene is trilayer. [24] (d) Image showing the backside of the silicon chip. (e) The optical image shows the location of Dev-1 and Dev-2 relative to the diaphragm center (red dot).

The schematic of the experiments is shown in Figure 2. The pressure on the device side of the diaphragm ($P_A$), referred to as the front side, is always kept at about $2 \times 10^{-7} kPa$. We vary the pressure on the backside ($P_B$) of the silicon diaphragm using a separate pressure/vacuum pump. The atmospheric pressure ($P_{atm}$) is assumed to be 91.6kPa [25]. The total range of pressure change that can be created using the pressure and vacuum pumps is from 20kPa to 270kPa. The effective pressure difference across the chip membrane is $\Delta P = P_B - 2 \times 10^{-7} kPa \approx P_B$. We connect the pressure or the vacuum pump to the back side of the package using the CF-flange to Ferrule converter and the necessary set of accessories.

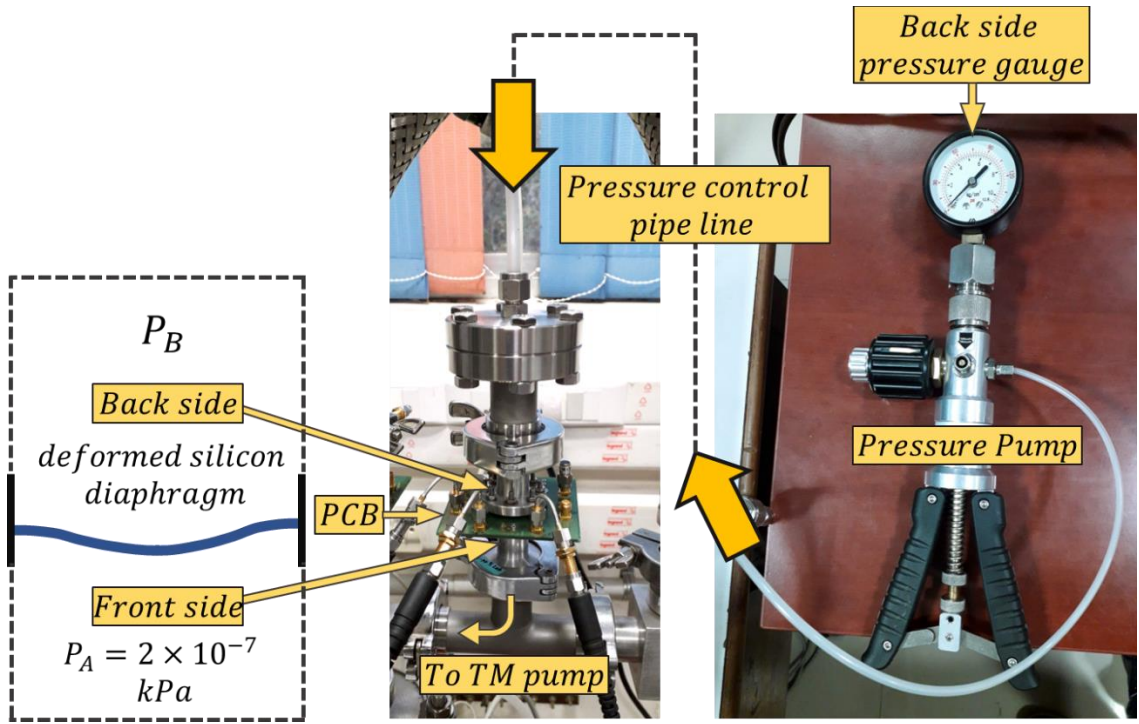

Figure 2: Schematic showing the experimental setup. The pressure on the back side of the PCB ($P_B$) is varied using a pressure/vacuum pump. The pressure gauge attached to the pump is used to monitor $P_B$. The front side pressure ($P_A$) is fixed at $2 \times 10^{-7} kPa$. The inset on left represents the simplified view of the deformed silicon diaphragm as a result of the positive pressure difference.

The electromechanical characterization of the device is done using the FM demodulation scheme [26]. The basic characterization of the resonance is done when the back of the silicon diaphragm is exposed to atmospheric pressure (i.e., $\Delta P = 91.6 kPa$). Figure 3(a and b) shows the resonance peak of the device for Dev-1 and Dev-2, respectively. A fitting routine [26] is used to estimate the device's resonance frequency and quality factor. Figure 3 (c) and (d) show the dispersion of the fundamental mode of Dev-1 and Dev-2 with DC gate voltage. The intrinsic strain is estimated using the following frequency-strain relation [27]

$$f_0 = \frac{1}{2\pi}\sqrt{\frac{\beta + 3\alpha z^2}{m}} \qquad (1)$$

where,

$$\beta = \frac{16\,E\,S\,\epsilon_0}{3L} - \frac{1}{2}C''V_g^2 \qquad (2)$$

$$\alpha = \frac{256\,E\,S}{9\,L^3} \qquad (3)$$

$$z = -\frac{0.87\beta}{\sqrt[3]{9\alpha^2\gamma + 1.7\sqrt{4\alpha^3\beta^3 + 27\alpha^4\gamma^2}}} + \sqrt[3]{9\alpha^2\gamma + 1.7\sqrt{4\alpha^3\beta^3 + 27\alpha^4\gamma^2}}\,\frac{1}{2.6\alpha} \qquad (4)$$

$$\gamma = -\frac{1}{2}C'V_g^2 \qquad (5)$$

$\epsilon_0$ is the strain at the 91.6kPa pressure difference, S is the surface area, L is the length, and m is the mass, E is Young's Modulus of the vibrating element. The length and width of the devices are obtained from the SEM images.

To separate the electrostatic effect from the built-in strain, we look at equation (1) at $V_g = 0$ (section S2 supplementary material). This is a good assumption because the pressure sensing mechanism relies on a change in the built-in strain as a result of the deformation of the diaphragm.

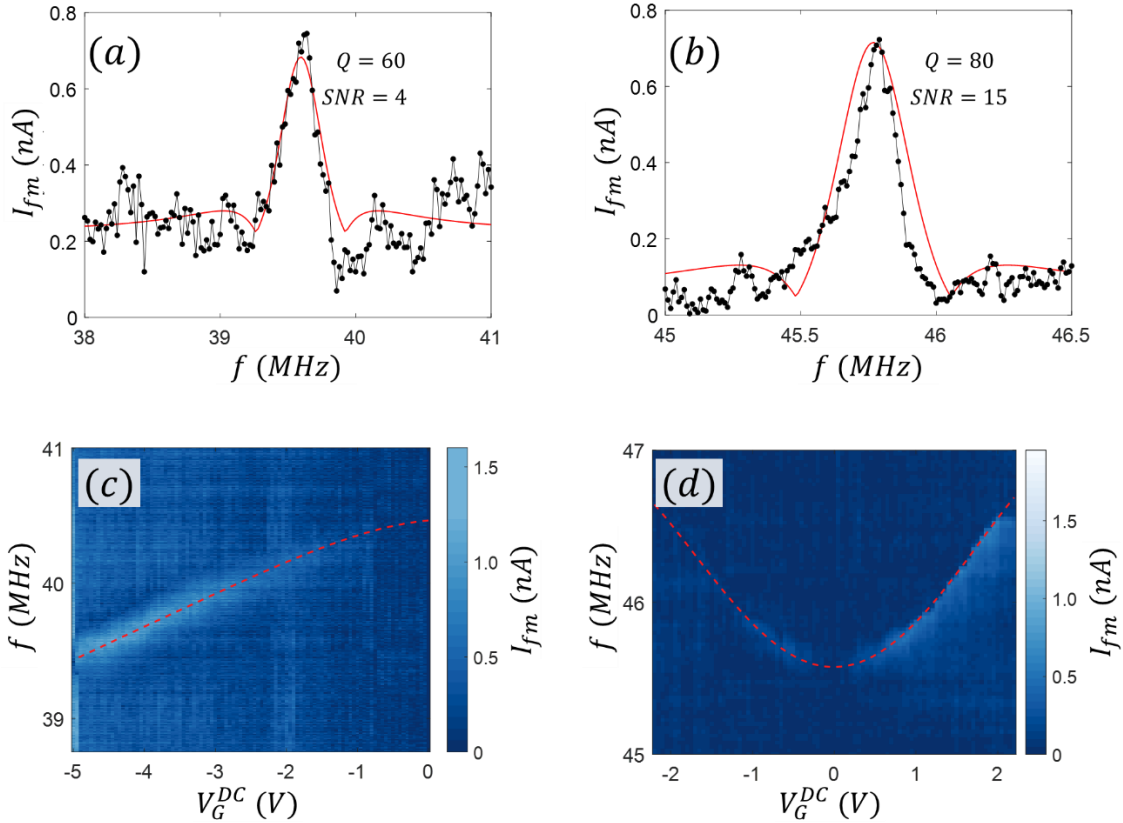

Figure 3: (a) and (b) the frequency response, and (c) and (d) show the mode dispersion Dev-1 and Dev-2, respectively. The frequency response is obtained at $V_s^{AC} = 350mV$, $V_G^{DC} = -5V$ for Dev-1 shown in (a) and $V_s^{AC} = 350mV$, $V_G^{DC} = 1V$ for Dev-2 in (b). The $V_s^{AC}$ for frequency dispersion measurements is $450mV$ and $350mV$ for Dev-1 and Dev-2 respectively. The quality factor and the SNR are extracted by fitting the resonance data. The dispersion of Dev-1 is recorded only for the negative DC gate voltages. The dispersion of Dev-1 shows an electrostatic softening due to high built-in strain. Dev-2 shows a hardening of resonance with DC gate voltage which results from the low built-in strain.

The pressure sensing experiments are performed in an open-loop configuration. The pressure on the back side ($P_B$) is set to the desired value, and a frequency sweep is performed to locate the resonance frequency. The dial on the hand pump reads the pressure on the backside of the diaphragm. One set of the experiment consists of changing the $P_B$ near atmospheric pressure in steps and recording the resonance frequency at each step. Figure 4 shows the data from the pressure sensing experiments on Dev-1 and Dev-2, respectively. For Dev-1, the pressure ($P_B$) is reduced from atmospheric pressure ($P_{atm} = 91.6 kPa$) to 40 kPa using the vacuum pump (Figure 4, a) and increased to 160 kPa using the pressure pump (Figure 4,b). Dev-2's corresponding limits are 20 kPa (Figure 4,c) and 270 kPa (Figure 4,d). The solid blue line shows the frequency change when $P_B$ is changed from the atmospheric pressure, and the dashed blue line shows the frequency change when $P_B$ returns back to the atmospheric pressure. The red line is a linear fit to the data. The absence of hysteresis in the two lines underscores the reversibility of the sensing scheme. Also evident is the linearity of the device response. The slope of the line gives the responsivity ($\mathcal{R} = \Delta f / \Delta p$) of $17\ Hz/Pa$ and $20\ Hz/Pa$ for Dev-1 and Dev-2 respectively.

Figure 5(a) shows the plot of equation (1) for Dev-1 in red and Dev-2 in blue. The frequency of the fundamental mode of Dev-1 and Dev-2, when the backside of the diaphragm is exposed to the atmosphere (i.e., $\Delta P = 91.6 kPa$), is shown by the red and blue dot, respectively. These points represent the device's state at the start of experiments. The slope, $\Delta f / \Delta \epsilon$, at the respective starting points, is used to approximate the frequency tuning of each device. It can be seen from this plot that the built-in strain in Dev-2 is lower than the built-in strain in Dev-1. The value of $\Delta f / \Delta \epsilon$ is 90GHz and 150GHz for Dev-1 and Dev-2, respectively. The built-in strain ($\epsilon_0$) is $2.7 \times 10^{-4}$ and $1.5 \times 10^{-4}$ for Dev-1 and Dev-2, respectively. This information is listed in Table 1 for quick reference. Despite being longer, Dev1 has $f_0$ comparable to Dev-2, indicating the high built-in strain. At high built-in strain, the slope $df/d\epsilon$ is small. This is the reason for the slightly less responsivity of Dev-1 (17 Hz/Pa) than Dev-2 (20 Hz/Pa).

Table 1: Dimensions, resonance frequency and built-in strain for Dev-1 and Dev-2

|  | **Dev-1** | **Dev-2** |
|---|---|---|
| Length ($\mu m$) | 2.9 | 2 |
| Width ($\mu m$) | 0.5 | 0.5 |
| $f_0 (MHz, at\ V_G^{DV} = 0)$ | 40.50 | 45.5 |
| Built-in strain ($\epsilon_0$) | $2.7 \times 10^{-4}$ | $1.5 \times 10^{-4}$ |

| $\frac{\Delta f_0}{\Delta \epsilon}$ (GHz) | 90 | 150 |
| --- | --- | --- |
| $\frac{\Delta f_0}{\Delta p}$ (Hz/Pa) | 17 | 20 |

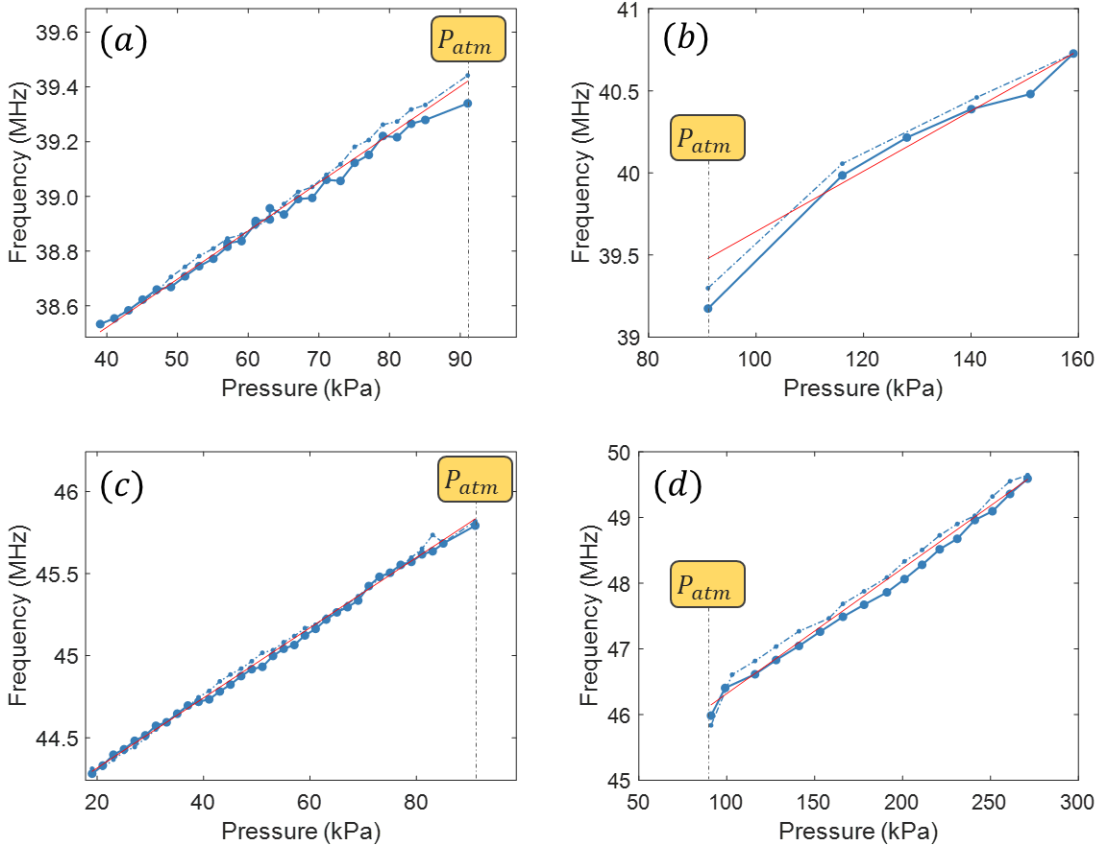

Figure 4: Frequency vs. Pressure plot for Dev-1 (a) and (b) and Dev-2 (c) and (d), The solid blue line with dots shows the frequency change for pressure change away from the $P_{atm}$ and the dashed blue line with dots shows the frequency change during reverse pressure change towards $P_{atm}$. The red line is a linear fit to the data.

We use solid mechanics model of the diaphragm to estimate strain change at the given location on the diaphragm when the diaphragm is subjected to a given pressure change, which is done easily by employing a finite element solver (COMSOL in our case). The $\Delta \epsilon$ vs $\Delta P$ using the FEM solver is plotted in Figure 5 (b) using a dashed-dotted line. Dev-1 and Dev-2 are at slightly different radial locations but close to the diaphragm center ($radial\ location \approx 0$). Given the $radius^2$ dependence of the strain, the strain change at the two locations close to the diaphragm center is not noticeably different. The red and blue lines show the strain change experienced by Dev-1 and Dev-2, respectively, based on the recorded frequency change for the given pressure change. The simulation overestimates the strain change by a factor of around 1.8 for Dev-1 and 2.5 for Dev-2. This error may arise due to the simplified solid mechanics model employed for the diaphragm and the assumption that the resonator experiences the full strain of the substrate.

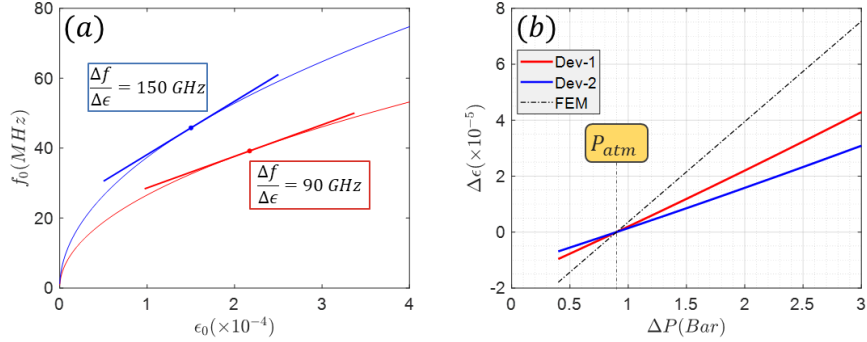

Figure 5: (a) Plot of strain vs. resonance frequency for Dev-1 (red) and Dev-2 (blue). The blue and red dots represent the frequency of the respective devices. The tangent to $\epsilon_0$ $vs$ $f_0$ plot at these points is used to approximate the frequency tuning. (b) Graph comparing the actual strain change experienced by the 2D NEMS and the simulated strain change of the diaphragm when subjected to varying pressure. The dash-dotted line shows the simulated strain change for both Dev-1 and Dev-2. Although Dev-1 and Dev-2 are at different radial locations, the strain change experienced by both devices is almost identical as they are close to the center of the diaphragm.

We estimate the minimum detectable pressure change using the experimentally expected value of Allan deviation for the graphene nanoresonator. First, we calculate the theoretical minimum of Allan deviation using [28]

$$\sigma_{th} \approx \frac{\sqrt{\langle Z_{th}^2 \rangle}}{2QA_c \sqrt{B2\pi\tau_A}} \tag{6}$$

For the Dev-2 $\sqrt{\langle Z_{th}^2 \rangle} = \sqrt{\frac{4Qk_bTB}{m\omega_0^3}} = 9.8 \times 10^{-14} m,$

$$\sigma_{th} \approx 6.8 \times 10^{-8} \quad (7)$$

With $Q = 80$, $T = 300K$, $B = 0.8Hz$, $\tau_A$=10s, $\omega_0$=2π 46× $10^6 \frac{rad}{s}$. The critical amplitude is assumed to be equal to the thickness of the trilayer graphene, $A_c = 3 \times 0.3nm$. This is a reasonable assumption to estimate the upper limit [29].

With the empirical rule that the actual frequency fluctuations are about 2.7 orders of magnitude greater than the theoretical Allan deviation [28], the actual Allan deviation for our measurement setup

$$\sigma_{ex} \approx 3.4 \times 10^{-5} \quad (8)$$

With frequency fluctuations $\Delta f = \sigma_{ex} f_0$, the minimum detectable pressure (resolution) with our setup is given by

$$\Delta P_{min} = \frac{\sigma_{ex} f_0}{\Delta f / \Delta P} \quad (9)$$

Using the pressure responsivity of $\frac{\Delta f}{\Delta P} = 20 \frac{Hz}{Pa}$, and $f_0 = 46 \, MHz$ the limit of detection is given by the following relation

$$\Delta P_{min} = \frac{\sigma_{ex} f_0}{\Delta f / \Delta P} \approx 90 Pa \quad (10)$$

Experimentally, this is the minimum change in pressure that can be detected using this device as a pressure sensor. Higher responsivity ($\Delta f / \Delta p$) is desirable for better resolution. The responsivity is a function of the location and orientation of the resonator on a given diaphragm [23] and the stiffness of the diaphragm itself. Different locations on the diaphragm offer different responsivity (see section S3 of SI). The strain change at the center is maximum and varies quadratically as a function of radial location. Assuming the pressure acts on the side opposite to the device side (i.e., back side), the radial component of strain is maximum at the center, and it is tensile. The radial strain reduces quadratically with radius (equation S2), becomes zero at a radius equal to $\frac{R_d}{\sqrt{3}}$, where $R_d$ is the radius of the diaphragm (S3), and then changes sign to become compressive maximum at the periphery. The compressive maximum radial strain is twice the magnitude of the tensile strain (S3). Assuming the responsivity corresponding to maximum tensile radial strain to be $\mathcal{R}_{max}$, thus responsivities ranging from $-2\mathcal{R}_{max}$ to $\mathcal{R}_{max}$ can be attained. This can be done for a given diaphragm by placing the resonator at an appropriate location on the diaphragm. Since the resonators are not exactly at the center, the responsivity is 0.2% less than the maximum achievable responsivity by placing the resonators at the center (S4), i.e., the experimentally observed responsivity of 20 Hz/Pa is close to the $\mathcal{R}_{max}$.

The responsivity can also be tuned by changing the dimensions of the silicon diaphragm. The responsivity can be expressed as the product of the rate of change of the diaphragm's strain with pressure and rate of change of frequency with strain, assuming that the strain change of the diaphragm is completely transferred to the nanoresonator,

$$\frac{\Delta f}{\Delta p} = \left(\frac{\Delta \epsilon}{\Delta p}\right)_{diaphragm} \times \left(\frac{\Delta f}{\Delta \epsilon}\right)_{nanoresonator} \quad (11)$$

Thus, the different responsivities can be accessed by changing the dimensions of the diaphragm. The diaphragm can be made thinner and wider to produce large strain change with small pressure change, thus improving responsivity. $\frac{\Delta \epsilon}{\Delta p}$ can be simulated for different aspect ratios of the diaphragm using commercial finite element solvers. For a diaphragm of $50 \mu m$ thickness and $4 mm$ diameter, the responsivity can be increased till $500\ Hz/Pa$, with a bursting pressure of $1.2\ Bars$, resulting in $\Delta p_{min} = 3.2\ Pa$ (Section S5 of SI). With increased responsivity, the range decreases commensurately. It can be seen from the responsivity and range product expressed as

$$\frac{\Delta f}{\Delta p} \times \Delta P_{max} = \left(\frac{\Delta f}{\Delta \epsilon}\right)_{nanoresonantor} \times \epsilon_{si}^{break} \quad (12)$$

Which is constant for the given nanoresonator design. For Dev-2, this number is $60 \times 10^6 \frac{Hz}{Pa} Pa$. The responsivity and range product gives the sense of maximum accessible responsivity (= 60MHz) for 1 $Pa$ range or a maximum accessible bursting pressure (60 MPa) for 1 $\frac{Hz}{Pa}$ responsivity.

Although we limited the maximum pressure to 270kPa, the bursting pressure of the silicon diaphragm sets the maximum pressure that can be detected. We estimate the maximum working pressure of 3000 kPa. This gives us a range of about 33000 times the minimum detectable pressure change. In other words, a pressure measurement's sensitivity (or accuracy) is 0.003% of the full-scale range (Section S5 of SI, equation S5).

The experimentally observed responsivity of Dev-2, 20Hz/Pa, is the highest among the reported resonant pressure sensors. Almost all the reported resonant pressure sensors use silicon resonators of various designs. The highest responsivity amongst them is about 300Hz/kPa [12]. Our design has a responsivity that is two orders higher than previously reported. The primary reason is the higher Young's modulus of graphene as compared to that of silicon. We compare the responsivity of Dev-2 with a few representatives of the reported resonant pressure sensors in Table 2. It is worth mentioning that there are resonant pressure sensors that use the squeeze film effect instead of strain tuning of the resonance. Even in this class of pressure sensors, graphene-based pressure sensors outperform silicon-based pressure sensors [20]. The highest reported graphene resonant pressure utilizing the squeeze film effect is 90Hz/Pa [20]. But this high responsivity is accessible only for a short range of pressure change from 10kPa to 50kPa (absolute). Beyond this range, the responsivity degrades, and the pressure vs. frequency response becomes nonlinear. Another disadvantage of resonant pressure sensors utilizing the squeeze film effect is that the resonating beam is exposed to the fluid exerting pressure, degrading the quality factor. Consequently, such pressure sensors find applications for low-pressure (vacuum) sensing. Our pressure sensor has the highest responsivity and a wide pressure range of up to 3000 kPa (absolute). In our experiments, we limited the maximum pressure to 270kPa. Thus, we utilized only 9% of the entire range, giving the range-responsivity product of 5.4MHz. The range-responsivity product gives a better metric for comparing resonant pressure sensor designs. We denote this product as RR in Table 2. Our device offers the highest range-responsivity number than other reported values.

Table 2: Comparison of performance of different pressure sensors with our pressure sensor

| Resonator type and reference | Resonance tuning Mechanism | Responsivity (Hz/kPa) | Range (kPa) | RR (MHz) | Quality factor |
|---|---|---|---|---|---|
| Graphene beam (our device) | Strain | $20 \times 10^3$ | 20-270 | 5.4 | 100 |
| Silicon structure [12] | Strain | 300 | 110-160 | 0.048 | $10^4$ |
| Dual silicon beams [14] | Strain | 166 | 50-110 | 0.018 | $11 \times 10^3$ |
| Silicon beam [13] | Strain | 140 | 15-130 | 0.018 | $10^5$ |
| Graphene membrane [20] | Squeeze film effect | $90 \times 10^3$ | 10-50 | 4.5 | 10 |
| Silicon Beam [30] | Squeeze film effect | $12 \times 10^3$ | 0.1-1 | 0.012 | $35 \times 10^3$ |

In conclusion, we have demonstrated resonant pressure sensing using the trilayer graphene nanoresonators. The nanoresonator's frequency response to the strain change of the thin $Si/SiO_2$ diaphragm is caused by the pressure difference. The devices show responsivity that is two orders of magnitude higher than the conventional silicon resonant pressure sensors. The high Young's modulus of graphene allows access to such high levels of responsivity. The proposed design of the resonant pressure sensors offers simplicity and practical convenience and paves the way for utilizing 2D materials-based NEMS for various resonant sensing schemes. Our research highlights the potential of graphene-based nanoresonators for various applications in pressure sensing and beyond. Further investigations could explore the optimization of the device design, fabrication techniques, and integration with complementary metal-oxide-semiconductor (CMOS) processes. This could pave the way for the large-scale utilization and commercialization of graphene-based resonant pressure sensors, enabling advancements in fields such as biomedical monitoring, environmental sensing, and industrial process control.

**Supplementary Material**

Supplementary material includes details about the sensor package and information about various calculations performed in the article.

**Acknowledgment**

We acknowledge funding support from MHRD, MeitY and DST Nano Mission through NNetRA.

# Supplementary Material: Graphene Resonant Pressure Sensor with Ultrahigh Responsivity

*Swapnil More and Akshay Naik*

*Centre for Nano Science and Engineering, Indian Institute of Science, Bangalore 560012*

**S1. The pressure sensor package details**

The PCB has a hole in the center (Figure S6). The device chip is bonded onto the PCB using an epoxy adhesive (Araldite™) such that the chip entirely covers the hole. The diameter of the hole in the PCB is 6mm, and the chip size is 9mm× 9$mm$. The PCB and the chip bonded onto it define the plane across which the pressure differential is created. Any pressure difference across the PCB also acts on the chip as the etched backside of the chip is accessible through the hole in PCB. Figure S6 shows the front and backside of the PCB. The backside of the diaphragm is visible from the hole on the backside of the PCB.

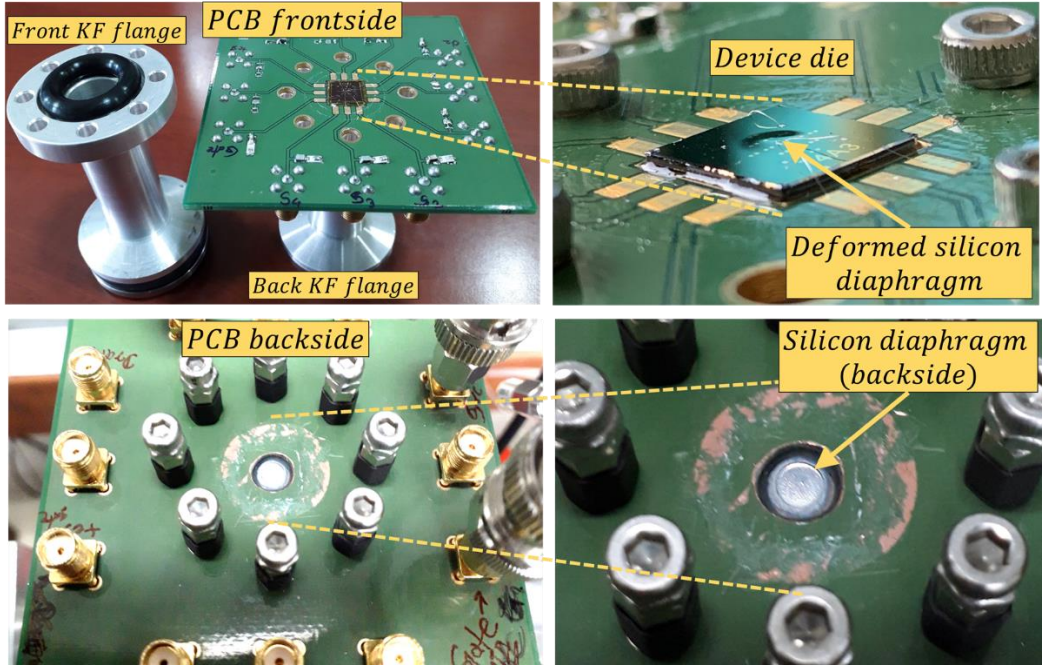

Figure S6. Images showing details of the PCB chip assembly from the front and back sides.

## S2. Effect of built-in strain on frequency dispersion of graphene nanoresonator

Equation (4) is the expression of the static displacement of the 2D membrane. The built-in strain and the gate voltage ($V_g$) control the resonant frequency of the graphene nanoresonator. The resonant frequency increases with increasing built-in strain and vice versa. The behavior of the frequency dispersion ($f_0$ $vs$ $V_G^{DC}$) changes with changing built-in strain ($\epsilon_0$). At small built-in strain, the increase in $V_g$ causes an increase in static displacement ($z$), resulting in an increase in resonance frequency. In contrast, at large built-in strain, the static displacement does not change much with increasing $V_g$. The electrostatic softening becomes much larger than the hardening due to stretching caused by electrostatic force. As a result, the resonance frequency reduces with increasing $V_g$. This is visible in Figures 3 (c) and (d). For Dev-1, built-in strain is higher than Dev-2; thus, electrostatic softening is dominant in Dev-1, whereas for Dev-2, hardening due to stretching caused by electrostatic forces is dominant.

## S3. Responsivity at different locations on the diaphragm

Pressure responsivity is defined as

$$\frac{\Delta f}{\Delta p} = \left(\frac{\Delta \epsilon}{\Delta p}\right)_{diaphragm} \times \left(\frac{\Delta f}{\Delta \epsilon}\right)_{nanoresonator} \qquad (S1)$$

The strain change per unit pressure change, $\left(\frac{\Delta \epsilon}{\Delta p}\right)_{diaphragm}$, is a function of the location of the resonator on the diaphragm. The strain change along the resonator aligned in the radial direction at radius $r$ on a diaphragm of radius $R_D$ and thickness $h$ subjected to pressure $P$ acting upwards on the bottom surface is given as [1]:

$$\epsilon_r = \left(\frac{-3PR_D(1-\nu^2)}{4Eh^2}\right)\left(\frac{3r^2}{R_D^2} - 1\right) \qquad (S2)$$

$E$ and $\nu$ are Young's modulus and poisons ratio of the diaphragm material. $\epsilon_r$ is maximum at $r = 0$. At $r = R_D/\sqrt{3}$ the radial strain becomes zero. Beyond this radius, the strain is compressive.

The ratio of strain at the center and the periphery is obtained as follows

$$\frac{\epsilon_r(r=0)}{\epsilon_r(r=R_D)} = \frac{-1}{3-1} = -\frac{1}{2} \qquad (S3)$$

Suppose there are two devices of the same dimensions and built-in strain, i.e. Δf/Δϵ is the same for both, one at the center and another at the edge of the diaphragm and both are aligned radially. Let $\mathcal{R}_{max}$ denote the responsivity of the device at the center. The responsivity of the device at the edge is found by using equation (S3) is $-2\mathcal{R}_{max}$.

We simulate the strain change for a silicon diaphragm of $150\mu m$ thickness, $2mm$ radius subjected to a pressure change of 100kPa. The strain is recorded on the top surface, while the pressure acts upwards on the back surface of the diaphragm. The simulation result is shown in the Figure S7. The figure shows radial as well as tangential components of strain change. We are interested in the radial component of strain. The radial component is maximum at the center ($r = 0$), and it is tensile. The $\Delta\epsilon_r$ reduces quadratically and becomes zero at $r = 2/\sqrt{3}$, as predicted by equation (S2). For $r > 2/\sqrt{3}$ the strain change is compressive. The compressive strain is maximum at the edge of the diaphragm ($r = 2$), and it is twice in the magnitude of the strain at the center of the diaphragm. From equation (S1), we can see that a similar resonator at $r = 0mm, 2/\sqrt{3}\ mm$ and $2mm$ will have responsivities (in $Hz/Pa$) $\mathcal{R}_{max}, 0$ and $-2\mathcal{R}_{max}$ respectively, where $\mathcal{R}_{max}$ is the responsivity of the deivce at the centre of the diaphragm, which experiences maximum tensile strain.

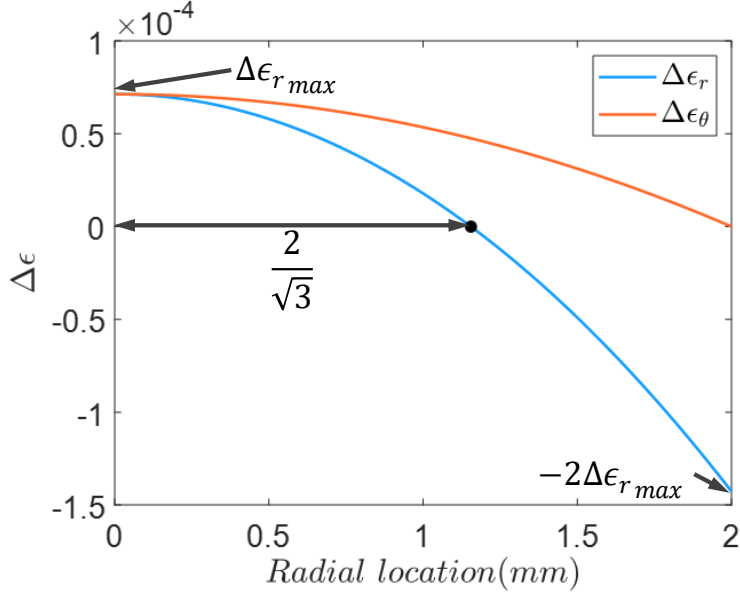

Figure S7: The simulated strain change for a $150\mu m$ thick, 2mm in radius silicon diaphragm subjected to 100kPa pressure difference.

**S4. Reduction in responsivity of Dev2.**

Dev2 is at a radial distance of $50\mu m$ from the diaphragm center. Using equation (S2), we can find the reduction in responsivity of Dev2

$$\frac{\epsilon_r(r=0)}{\epsilon_r(r=50\times 10^{-6})} = \frac{-1}{3\left(\frac{50\mu m}{2mm}\right)^2 - 1} = 1.0019 \qquad (S4)$$

Thus, for a given pressure change, the strain at $r = 50\mu m$ is roughly 0.2% lower than at the center. Thus, by placing a device at $50\mu m$ away from the center, the responsivity reduces by 0.2%.

**S5. Effect of diaphragm geometry responsivity, $\Delta P_{max}$, and $\Delta P_{min}$**

We assume the breaking strain of the silicon diaphragm to be $4 \times 10^{-4}$ (Considering the factor of safety of 2.5 and ultimate breaking strain of 0.1% for silicon [2]). The bursting pressure is the pressure required to increase the strain in the silicon diaphragm to $4 \times 10^{-4}$, which is given by

$$\Delta P_{max} = \frac{\Delta p}{\Delta \epsilon} \times \epsilon_{si}^{break} = \frac{\Delta f / \Delta \epsilon}{\Delta f / \Delta p} \times \epsilon_{si}^{break} = \frac{150 \times 10^9}{20} \times 4 \times 10^{-4} = 30 \times 10^5 Pa \qquad (S5)$$

Equation (12) in the main text, reproduced here as equation (S6), provides a design rule for the diaphragm to achieve the required responsivity. The equation relates the responsivity to the geometry of the diaphragm and stiffness of the nanoresonator

$$\frac{\Delta f}{\Delta p} = \left(\frac{\Delta \epsilon}{\Delta p}\right)_{diaphragm} \times \left(\frac{\Delta f}{\Delta \epsilon}\right)_{nanoresonator} \qquad (S6)$$

The term, $\frac{\Delta f}{\Delta \epsilon}$, is the property of the graphene resonator (Figure 5), $\frac{\Delta f}{\Delta \epsilon}$ is assumed to be 150 GHz, and the second term, $\frac{\Delta \epsilon}{\Delta p}$, depends on the compliance of the diaphragm. $\frac{\Delta \epsilon}{\Delta p}$ is the strain change produced for the given pressure change. A compliant diaphragm can achieve higher strain change for a given pressure change. Thus, by modifying the dimensions of the silicon diaphragm, the value $\frac{\Delta \epsilon}{\Delta p}$ can be changed. And by selecting the appropriate dimensions of the silicon diaphragm, we can design a pressure sensor with the required responsivity and range of operation. To estimate responsivity as a function of diaphragm geometry, the $\frac{\Delta \epsilon}{\Delta p}$ is calculated for a circular diaphragm for a range of thickness and diameter using the FEM tool (Figure S8 (a)). The simulated $\frac{\Delta \epsilon}{\Delta p}$ is used to calculate the responsivity ($\Delta f / \Delta p$), $P_{max}$, and $P_{min}$ as a function of diaphragm geometry using equations 12, 11 and 9 in the main text, respectively. The simulation is performed for a radius from 0.5 mm to 4mm and the diaphragm thickness from 2 μm to 100 μm. Note that a thinner, wider diaphragm is more compliant than a thicker, smaller diaphragm. Change in strain for a given change in pressure is more for a compliant diaphragm than a stiffer diaphragm. Thus to improve responsivity, a thinner and wider diaphragm is better than a stiffer diaphragm. For our current design of the diaphragm, d=4mm, t=150μm, the Δϵ/Δp ≈ 1.33×10^(-5) Bar^(-1). If we reduce the thickness of the diaphragm to 50μm, the Δϵ/Δp improves to 3.2×10^(-4) Bar^(-1), giving the improved responsivity of 480 Hz/Pa. In these calculations $\Delta f/\Delta \epsilon$ is assumed to be 150GHz.

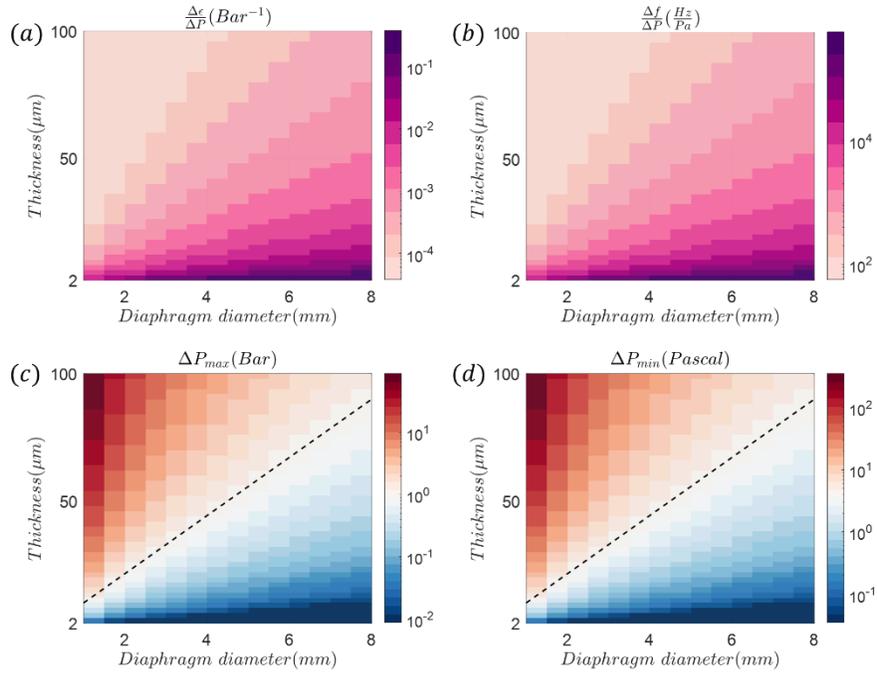

Figure S8: (a) Simulated values of Δϵ/ΔP, (b) Δf/ΔP, (c) $\Delta P_{max}$, (d) $\Delta P_{min}$, for various diaphragm sizes. For a thinner (or wide diameter) diaphragm, Δϵ/ΔP is large compared to a thicker (or small diameter) diaphragm. Therefore the responsivity (Δf/ΔP) is higher, and $\Delta P_{max}$ and $\Delta P_{min}$ are lower for a thinner or wider diaphragm,

As the diaphragm's thickness increases or the diaphragm's diameter reduces, the $\Delta P_{max}$ increases as a result of a stiffer diaphragm. The $\Delta P_{max}$ is also the safe operating pressure for the pressure sensor. Suppose the sensing diaphragm is to be exposed to atmospheric pressure. In that case, it must have a bursting pressure above atmospheric pressure (> 1 bar, assuming the other side of the diaphragm is in an ultra-high vacuum, which is true in our case). Thus $\Delta P_{max} \geq 1$bar marks the region on the design plane for sensing above atmospheric pressure. This boundary is marked by the dashed black line in Figure S8 (c-d). For Dev-1 and Dev-2, the diaphragm is 4mm in diameter and 150 μm thick. With $\Delta P_{max}$ =30bar and $\Delta P_{min}$ =90Pa. By reducing the diaphragm thickness to 50μm, the $\Delta P_{min}$ can be improved to 3.2Pa with $\Delta P_{max}$ reduced to 1.2 bar. With a limited range of 1.2 bar, devices on such a diaphragm can still be used for near atmospheric conditions with resolution improved by one order of magnitude, i.e., from 95Pa to 3.2Pa. Thus Figure S8 provides an important resource for selecting the appropriate dimension of the diaphragm based on resolution and operating range requirements.